\documentclass[12pt]{article}
\usepackage{graphicx}
\begin{document}

\author{Gediminas Gaigalas, Rasa Karpu\v skiene, Zenonas Rudzikas \\
{\em Institute of Theoretical Physics and Astronomy, A. Go\v{s}tauto 12, } \\
{\em Vilnius 2600, LITHUANIA e-mail: gaigalas@mserv.itpa.lt}}
\date{}
\title{Optimal classification of HCI spectra}
\maketitle


\begin{abstract}
Energy levels of highly charged ions
 as a rule cannot be classified using
$LS$ coupling due to rapid increase of
relativistic effects.
It is suggested,
for optimal classification of
energy spectra, to
calculate them
in $LS$ coupling and to
transform the weights of the wave functions, obtained
after diagonalization of the energy matrix, to the other coupling
schemes.  F-like ions are considered as an example.

\end{abstract}

\section{\bf Introduction}

For optimal classification of
the energy spectra measured, the collaboration of experimenters
and theoreticians is required, because only from calculations
it is possible to find the closest to reality sets of quantum numbers.
The best way to achieve this is found
by calculations
of energy spectra in $LS$ coupling and
transformation of the weights of the wave functions, obtained
after diagonalization of energy matrix, to the other
schemes~\cite{R}.

Let us consider the energy levels of the configuration
$1s^{2} 2s^{2} 2p^{4} 3d~J=5/2$ (F-like isoelectronic sequence,
Al~V - Ge~XXIV) as an example.
Let us number them, for Al~IV, $J=5/2$, as follows: $(^3P)~^4D_{5/2}$ - 1,
$(^3P)~^4F_{5/2}$ - 2, $(^3P)~^4P_{5/2}$ - 3, $(^3P)~^2F_{5/2}$ - 4,
$(^3P)~^2D_{5/2}$ - 5, $(^1D)~^2F_{5/2}$ - 6, $(^1D)~^2D_{5/2}$ - 7,
and $(^1S)~^2D_{5/2}$ - 8.
If, calculating the energy spectra, we start with most realistic
coupling scheme, then the intermediate coupling will not differ much from
the pure one. In such a case one weight will be much larger compared  to the
others
in the eigenfunction obtained after the diagonalization of the energy matrix.
Then the quantum numbers of this largest
weight are used to classify the relevant energy level. If there are several
almost equal weights, then the initial coupling
is not suitable
and we have to look for the other coupling schemes.
Indeed, e. g., if we look at the classification of the
energy levels in Fawcett \cite{F}, we shall see that there are cases when
the quantum numbers of the $LS$ coupling can be hardly used for this purpose.

Similar conclusions can be derived from the relevant theoretical studies.
Table I presents the percentage of the largest weights of the levels 2-5
for F-like ions in $LS$ coupling. It illustrates that with the increase
of the ionization degree the accuracy of the $LS$ coupling worsens.
There are even cases when it is impossible to ascribe the largest weights
to certain levels (e.g. levels 2 and 3 for Cr~XVI and Mn~XVII as well
as the levels 2 and 5 for Cu~XXI and Zn~XXII).
Therefore we need a fairly universal methodology of
the classification of the
levels of atoms and ions with the most reliable sets of quantum numbers.
This is of particular importance for the case of the highly ionized atoms.

\begin{center}
{\bf Table I.}
\hspace{0.2cm}
{\rm Percentage of largest weights of the {\bf levels 2- 5} for $F$-like ions
in the configuration $1s^{2} 2s^{2} 2p^{4} 3d$ in $LS$ coupling.}
\begin{tabular}{ l r r r r r }  \hline \hline
Element & $Z$ & ~~~~~level 2&~~~~~level 3&~~~~~level 4&~~~~~level 5 \\[0.1cm]
\hline
Al~V    &13&$^{4}F$~(92\%)&$^{4}P$~(63\%)&$^{2}F$~(63\%)&$^{2}D$~(90\%) \\
Si~VI   &14&$^{4}F$~(91\%)&$^{4}P$~(83\%)&$^{2}F$~(80\%)&$^{2}D$~(88\%) \\
P~~VII  &15&$^{4}F$~(90\%)&$^{4}P$~(85\%)&$^{2}F$~(82\%)&$^{2}D$~(85\%) \\
S~~VIII &16&$^{4}F$~(88\%)&$^{4}P$~(84\%)&$^{2}F$~(80\%)&$^{2}D$~(82\%) \\
Cl~IX   &17&$^{4}F$~(85\%)&$^{4}P$~(81\%)&$^{2}F$~(77\%)&$^{2}D$~(78\%) \\
Ar~X    &18&$^{4}F$~(81\%)&$^{4}P$~(76\%)&$^{2}F$~(76\%)&$^{2}D$~(72\%) \\
K~~XI   &19&$^{4}F$~(76\%)&$^{4}P$~(69\%)&$^{2}F$~(67\%)&$^{2}D$~(69\%) \\
Ca~XII  &20&$^{4}F$~(69\%)&$^{4}P$~(61\%)&$^{2}F$~(61\%)&$^{2}D$~(65\%) \\
Sc~XIII &21&$^{4}F$~(62\%)&$^{4}P$~(53\%)&$^{2}F$~(56\%)&$^{2}D$~(61\%) \\
Ti~XIV  &22&$^{4}F$~(54\%)&$^{4}P$~(44\%)&$^{2}F$~(51\%)&$^{2}D$~(57\%) \\
V~~XV   &23&$^{4}F$~(45\%)&$^{4}P$~(37\%)&$^{2}F$~(46\%)&$^{2}D$~(54\%) \\
Cr~XVI  &24&$^{4}F$~{\bf (37\%)}&$^{4}F$~{\bf (42\%)}
&$^{2}F$~(43\%)&$^{2}D$~(51\%) \\
Mn~XVII &25&$^{4}F$~{\bf (30\%)}&$^{4}F$~{\bf (47\%)}
&$^{4}P$~(40\%)&$^{2}D$~(48\%) \\
Fe~XVIII&26&$^{2}F$~(27\%)&$^{4}F$~(51\%)&$^{4}P$~(42\%)&$^{2}D$~(46\%) \\
Co~XIX  &27&$^{2}F$~(27\%)&$^{4}F$~(52\%)&$^{4}P$~(43\%)&$^{2}D$~(44\%) \\
Ni~XX   &28&$^{2}F$~(27\%)&$^{4}F$~(52\%)&$^{4}P$~(43\%)&$^{2}D$~(42\%) \\
Cu~XXI  &29&$^{2}D$~{\bf (27\%)}&$^{4}F$~(51\%)&$^{4}P~(43\%)$&
$^{2}D$~{\bf (39\%)} \\
Zn~XXII &30&$^{2}D$~{\bf (28\%)}&$^{4}F$~(48\%)&$^{4}P~(43\%)$&
$^{2}D$~{\bf (37\%)}        \\
Ga~XXIII&31&$^{2}D$~(29\%)&$^{4}F$~(45\%)&$^{4}P~(42\%)$&$^{2}F$~(36\%)  \\
Ge~XXIV &32&$^{2}D$~(29\%)&$^{4}F$~(41\%)&$^{4}P~(41\%)$&$^{2}F$~(39\%)  \\
   \hline \hline
\end{tabular}
\end{center}

\section{\bf Methodology}

In general we shall follow the methodology of the optimization of the
coupling scheme described in \cite{R,RC}. If we calculate the energy matrix in
$LS$ coupling and diagonalize it, then we arrive at the eigenfunction in the
form
\begin{equation}
\label{eq:a}
\begin{array}[b]{c}
\Psi (\beta J) = \displaystyle {\sum_{ \alpha_{i} L_{i} S_{i}}}
a(\alpha_{i} L_{i} S_{i} J) \Psi ( \alpha_{i} L_{i} S_{i}J),
\end{array}
\end{equation}
where $\Psi ( \alpha_{i} L_{i} S_{i}J)$ is the wave function in
the pure $LS$ coupling and $a(\alpha_{i} L_{i} S_{i} J)$ are the weights of the
wave functions of that  coupling.
Calculations show that for the level 3 of the V~XV
the wave function of the intermediate coupling has two practically
equal weights, namely 0.607 and 0.594. Similar situation is also for the
levels 3 - 5. Very close are the weights of the functions
$\Psi ( 2p^{4}~ (^{3}P)~ 3d~ ^{2}D_{5/2})$ and
$\Psi ( 2p^{4}~ (^{3}P)~ 3d~ ^{2}F_{5/2})$
of the level 2 for Cu~XXI, namely 0.518 and 0.514. Mn~XIV has two practically
equal weights for level 4, namely
$a ( 2p^{4}~ (^{3}P)~ 3d~ ^{2}F_{5/2})$=0.633 and
$a ( 2p^{4}~ (^{3}P)~ 3d~ ^{4}P_{5/2})$=0.635.

There exists the following relationship between the weights of the wave
functions
of two pure coupling schemes~\cite{R}:
\begin{equation}
\label{eq:a-b}
\begin{array}[b]{c}
a_{ij} = \displaystyle {\sum_{ k}}
c_{ik} (\Psi_{i} \mid \Phi_{k}).
\end{array}
\end{equation}
Here $(\Psi_{i} \mid \Phi_{k})$ stands for the transformation matrix from
one coupling scheme to the other. Their general
expressions as well as simplified formulas for special cases may be
also found in~\cite{R}. Thus, knowing the weights $c_{ik}$ of the wave
functions in a certain pure coupling scheme, let us say $LS$, we are able
to find them for the case of any other coupling and to choose in this way
the optimal one, avoiding the calculation and the diagonalization of the
energy matrices in other coupling schemes. The relevant procedure is very
simple, particularly when having the necessary computer code.
Let us illustrate its practical use for the
optimization of the classification of the energy levels 2 and 3
of the F-like ions.  The levels of the
configuration $2p^{4}$ are classified using $LS$ coupling whereas
four coupling schemes ($LS$, $LK$, $JK$ and $JJ$) are
possible for coupling the angular momenta of $2p^{4}$ electrons with those of
$3d$ electron.
The energy levels in the $LK$, $JK$ and $JJ$ coupling schemes may be denoted
as $L_{1}[K]_{J}$, $J_{1}[K]_{J}$ and $[J_{1} J_{2}]_{J_{1}}$ correspondingly.

\section{\bf Discussion}

It turned out that the level 3 for low
ionization degrees (Al~V - Ar~X) may be also classified with
$LK$ coupling. However $LK$ coupling is extremely unfit for Ca~XII - Fe~XVIII.
The $JK$ coupling is slightly preferable in the middle of the isoelectronic
sequence, whereas $JJ$ coupling at the end.
However, for the ionization degrees studied the
rather large deviations from all considered pure coupling schemes including
$LS$ are observed.

Mean deviation $R$ of the vectors {\bf C} of intermediate coupling
from pure coupling {\bf T} may be
also used as the measure of the applicability of coupling
scheme (see~\cite{RC}):
\begin{equation}
\label{eq:d}
\begin{array}[b]{c}
R = \frac{1}{n} \displaystyle {\sum_{ k=1}^{n}}
\sqrt{ \displaystyle {\sum_{ i=1}^{n}}
(c_{ik}-t_{ij_{k}})^{2}}
= \frac{1}{n} \displaystyle {\sum_{ k=1}^{n}} \sqrt{2(1-c_{j_{k}k})}.
\end{array}
\end{equation}
Second equality in (\ref{eq:d}) is obtained making use of the orthonormality
conditions of the vectors {\bf C} and {\bf T} considered in the basis
of the same
coupling scheme i.e., when $t_{ij}= \delta_{ij}$.
Here $j_{k}$ denotes the largest component of the $k$-th vector and $n=4$.
Using the codes worked out let us calculate $R$
for the levels mentioned above.
Considering the weights of separate levels we see that the
validity of one or another coupling scheme varies
from one level to the other. On the contrary, $R$ is an
averaged quantity. It reflects the main features of the whole group of the
levels.

\begin{figure}
\centering
\hspace{-5.cm}
\includegraphics[height=15.0cm]{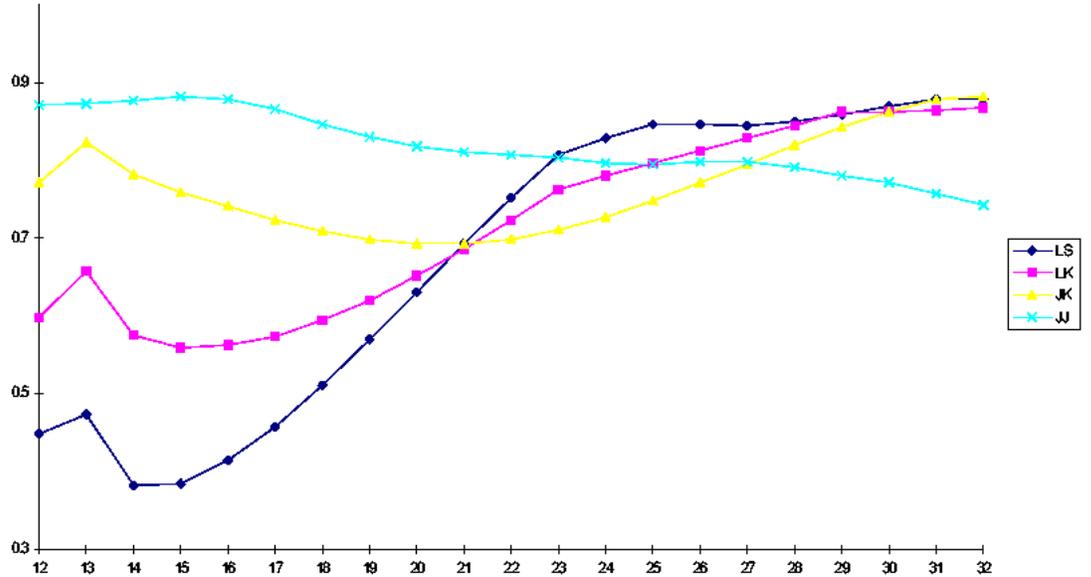}
\vspace{-5.cm}
\caption{Z dependence of $R$ for levels 2 - 5.}
\label{fig:e-b}
\end{figure}

Figure 1 illustrates the dependence of $R$ on
$Z$ for the levels 2-5. Dotted line and curves with filled squares,
triangles and circles respectively correspond to the $LS$, $LK$, $JK$ and
$JJ$ coupling schemes.
We see from Figure 1 that the dependence of $R$ on $Z$ may be devided
into three intervals. In the first interval $Z$=12 - 20 $LS$ coupling is the
best. Moreover, at the beginning of the
isoelectronic sequence ($Z$ = 12 - 17) it is valid with high accuracy
($R < $ 0.5) but with the increase of the ionization degree its quality
worsens. In this interval $LK$ coupling occupies the second place, but
with the increase of $Z$ it also worsens whereas the quality of $JK$
coupling improves. Around Sc~XIII ion all three coupling schemes ($LS$, $LK$
and $JK$) are practically equally valid (their $R$ values are respectively
equal to 0.693, 0.686 and 0.692).
In the second interval ($Z$ = 21 - 26) the $JK$ coupling is the most suitable.
Here $R$ value is decreasing (with the increase of $Z$)
only for $JJ$ coupling.
In the third interval ($Z$ = 28 - 32) the $JJ$ coupling is the best.
Only for it the $R$ value is decreasing with the increase of $Z$,
whereas for the rest three coupling schemes $R$ is practically the
same and at the end ($Z$ = 31, 32) it reaches 0.880.

\section{\bf Conclusion}

The levels 1, 6 - 8 for Al~V - Ge~XXIV in the configuration
$1s^{2}2s^{2}2p^{4}3d$
may be fairly accurately identified
with the help of $LS$ coupling.
However, the identification of the rest levels 2 - 5 based on the $LS$
coupling for some ionization stages is doubtful.

Considering the validity of the various coupling schemes for
levels 2 - 5, using the analysis of the structure of the weights of
the eigenfunctions or the quantity $R$, we arrive at similar conclusions.
In both cases we see that at the beginning of the sequence the
$LS$ coupling is valid with high accuracy
whereas at its end $JJ$ coupling prevails. Namely, the optimal classification of
the levels 2 - 5 would be: $LS$ for Al~V - Ca~XII, $JK$
for Sc~XII - Co~XIX and $JJ$ for the more highly ionized atoms.

The methodology of the optimization of the coupling scheme,
based on the transformation of the weights of a given pure coupling
scheme to other couplings and on the computer codes written, allow us to find
the optimal coupling scheme for classification of the energy levels of
atoms and ions, multiply charged ions included. The methodology and the
codes are fairly universal allowing a study of the
complex many-electron configurations. They may be particularly useful
when combined with relevant experimental studies.


%
%
%
%
%
%
%
%

\begin{thebibliography}{99}

\bibitem{R} Rudzikas, Z., "Theoretical Atomic Spectroscopy"
(Cambridge University Press, Cambridge 1997), p. 424.

\bibitem{F} Fawcett, B. C., Atomic Data and Nuclear Data Tables
{\bf 31}, 495 (1984).

\bibitem{RC} Rudzikas Z. and \v Ciplys, J., Physica Scripta
{\bf T26}, 21 (1989).

\end{thebibliography}
\end{document}